\documentclass[onecolumn, 11pt, aps, prd, amsmath, amsfonts, amssymb, floatfix, nofootinbib, notitlepage, preprintnumbers, superscriptaddress,qsymbols,epsf]{revtex4-1}
\pdfoutput=1

\usepackage{amsmath, amsfonts, amssymb}
\usepackage{color, graphicx, perpage}

\usepackage[english]{babel}
\usepackage[utf8]{inputenc}
\usepackage[T1]{fontenc}
\usepackage{ae}

\MakePerPage[2]{footnote}

\newcommand{\eq}[1]{Eq.~(\ref{#1})}

\newcommand{\fig}[1]{Fig.~\ref{#1}}

\renewcommand{\sec}[1]{Sec.~\ref{#1}}

\usepackage{soul}

\begin{document}

\preprint{IPMU16-0205}
\preprint{LPT-Orsay-16-83}

\title{\Large SU(5) Unification with TeV-scale Leptoquarks}

\author{Peter Cox}
\email{peter.cox@ipmu.jp}
\affiliation{Kavli Institute for the Physics and Mathematics of the Universe (WPI), UTIAS, The University of Tokyo, Kashiwa, Chiba 277-8583, Japan}

\author{Alexander Kusenko}
\email{kusenko@ucla.edu}
\affiliation{Department of Physics and Astronomy, University of California, Los Angeles, CA, 90095-1547, USA}
\affiliation{Kavli Institute for the Physics and Mathematics of the Universe (WPI), UTIAS, The University of Tokyo, Kashiwa, Chiba 277-8583, Japan}

\author{Olcyr Sumensari}
\email{olcyr.sumensari@th.u-psud.fr}
\affiliation{Laboratoire de Physique Th\'{e}orique, CNRS, Univ.Paris-Sud, Universit\'{e} Paris-Saclay, 91405 Orsay, France} 
\affiliation{Instituto de F\'{i}sica, Universidade de S\~{a}o Paulo, C.P. 66.318, 05315-970 S\~{a}o Paulo, Brazil}

\author{Tsutomu T.~Yanagida}
\email{tsutomu.tyanagida@ipmu.jp}
\affiliation{Kavli Institute for the Physics and Mathematics of the Universe (WPI), UTIAS, The University of Tokyo, Kashiwa, Chiba 277-8583, Japan}

%%%%%%%%%%%%%%%%%%%%%%%%%%%%%%%%%%%%%%%%

\begin{abstract}
\vspace{0.5cm}
\baselineskip=15pt
It was previously noted that SU(5) unification can be achieved via the simple addition of light scalar leptoquarks from two split $\bf10$ multiplets. We explore the parameter space of this model in detail and find that unification requires at least one leptoquark to have mass below $\approx16\,$TeV. We point out that introducing splitting of the $\bf24$ allows the unification scale to be raised beyond $10^{16}$ GeV, while a U(1)$_{PQ}$ symmetry can be imposed to forbid dangerous proton decay mediated by the light leptoquarks. The latest bounds from LHC searches are combined and we find that a leptoquark as light as 400 GeV is still permitted. Finally, we discuss the interesting possibility that the leptoquarks required for unification could also be responsible for the $2.6\sigma$ deviation observed in the ratio $R_K$ at LHCb.\end{abstract}

\maketitle

%%%%%%%%%%%%%%%%%%%%%%%%%%%%%%%%%%%%%%%%

\section{Introduction}

Grand unification has long provided a strong source of motivation when considering physics beyond the Standard Model. Despite its elegant simplicity, it is well-known that the original SU(5) model of Georgi and Glashow~\cite{Georgi:1974sy} is no longer viable. It suffers from several issues including: (i) doublet-triplet splitting;  (ii) Yukawa relations in disagreement with experiment; (iii) massless neutrinos; and perhaps most strikingly, (iv) fails to achieve unification. However, with the exception of the first problem, there are straightforward, non-supersymmetric extensions of the original model which can solve each of these issues. The fermion mass relations can be addressed using higher-dimension operators~\cite{Ellis:1979fg} or a Higgs in the $\bf45$ representation~\cite{Georgi:1979df}, while the addition of singlet right-handed neutrinos allows for neutrino masses via the Type-I seesaw~\cite{Minkowski:1977sc,*Yanagida:1979as,*Glashow:1979nm,*GellMann:1980vs}. Then there exist several models which introduce additional split multiplets in order to achieve gauge coupling unification consistent with the current experimental measurements~\cite{Murayama:1991ah,Giveon:1991zm,hep-ph/0504276,hep-ph/0612029,1604.03377}. 

In this paper, we revisit the model originally proposed in Ref.~\cite{Murayama:1991ah}, where it was demonstrated that the introduction of two ({\bf3},{\bf2},1/6) scalar leptoquarks and a second Higgs doublet at the electroweak scale can be used to achieve coupling unification. An up-to-date analysis of this model is particularly interesting in light of several recently observed anomalies in the decays of $B$-mesons. In particular, the lepton flavour universality violating ratio $R_K$ measured at the LHCb experiment, which shows a $2.6\sigma$ deviation with respect to the SM prediction. It was shown in Ref.~\cite{Becirevic:2015asa} that the anomaly can be explained\footnote{There have also been other proposed leptoquark explanations of the $R_K$ anomaly~\cite{Hiller:2014yaa,*Ghosh:2014awa,*Alonso:2015sja,*Fajfer:2015ycq,*Barbieri:2015yvd,*Deppisch:2016qqd,*Arnan:2016cpy,*Becirevic:2016yqi,*Hiller:2016kry,*Barbieri:2016las}.} by the addition of a scalar leptoquark transforming under the SM gauge group as ({\bf3},{\bf2},1/6). If the observed discrepancy is confirmed with more data, this measurement could therefore provide the first hints in favour of unification along the lines of that proposed in~\cite{Murayama:1991ah}.

However, the original model now faces several difficulties, including a low unification scale ($\sim5\times10^{14}\,$GeV) potentially in tension with the bounds from Super Kamiokande, as well as additional contributions to proton decay mediated by the light leptoquarks. Furthermore, the original assumption of leptoquarks with masses close to $m_Z$ has since been excluded by direct searches.

In this work we show that the model of Ref.~\cite{Murayama:1991ah} in fact still remains viable when confronted with the latest experimental constraints. Firstly, we point out that extending the model to introduce splitting in the $\bf24$ Higgs can preserve unification, while also allowing one to significantly raise the scale of unification. Furthermore, we discuss how dangerous contributions to proton decay, mediated by the light leptoquarks, can be forbidden by a U(1)$_{PQ}$ symmetry. We explore in detail the parameter space of the model and find that at least one of the leptoquarks is required to have a mass below $\lesssim16\,$TeV. Flavour constraints on such light leptoquarks are then discussed in \sec{sec:flavour}, in particular the intriguing possibility that the observed anomaly in $R_K$, should it persist, could provide the first evidence for the existence of such states. Finally, the lightest leptoquark could be within the reach of direct searches at the LHC. In \sec{sec:LHC} we combine the latest limits from LHC searches and find that a leptoquark as light as $\sim400\,$GeV is still allowed by the current data. 

%%%%%%%%%%%%%%%%%%%%%%%%%%%%%%%%%%%%%%%%

\section{Model} \label{sec:model}

We consider a simple extension of the Georgi-Glashow model that was first proposed in Ref.~\cite{Murayama:1991ah}. The scalar sector is extended to include an additional $\bf\overline{5}$ Higgs $H_{\overline{5}}$, as well as two new scalars, $\Phi^{(1)}_{10}$ and $\Phi^{(2)}_{10}$, transforming in the $\bf10$ representation. Experimental bounds on the proton lifetime provide a lower limit ($m_T\gtrsim10^{12}\,$GeV) on the masses of the colour-triplet Higgs in the $\bf5$ and $\bf\overline{5}$, leading to the well-known doublet-triplet splitting problem. These triplet Higgs are therefore assumed to acquire GUT scale masses. This motivates the assumption that such splitting could in fact be a generic feature of the scalar sector, which then opens new avenues to achieve unification. While this generically requires additional fine-tuning, naturalness is already assumed not to be a valid guiding principle within the context of non-supersymmetric GUTs.

The decouplets, $\Phi_{10}$, can be decomposed under the SM gauge group as
\begin{equation}
  {\bf10}=({\bf3},{\bf2},1/6) \oplus ({\bf\overline{3}},{\bf1},-2/3) \oplus ({\bf1},{\bf1},1) \,.
\end{equation}
Splitting of the $\bf10$ is assumed such that the $\Delta\equiv({\bf3},{\bf2},1/6)$ can remain light, while the rest of the multiplet acquires GUT scale masses. This is the key assumption that leads to unification in this model. 

Departing from the original model, we will also consider the case where there is a splitting of the $\Sigma_{24}$ Higgs by lowering the mass of the ({\bf8},{\bf1},0) and ({\bf1},{\bf3},0). Naively, one would expect this additional splitting to disrupt unification. However, if the octet and triplet are approximately degenerate in mass, then their combined effect on the RGEs is such that unification can be preserved. As we shall demonstrate in the following section, splitting of the $\bf24$-plet then provides a straightforward way to raise the unification scale. Such behaviour was first pointed out in the context of supersymmetric, string-motivated models in Ref.~\cite{hep-th/9510094}. In addition to the minimal case, we will consider the possibility that the $\bf24$-plet is described by a complex scalar field. The additional octet and triplet degrees of freedom then allow further raising of the unification scale. However, in the case where $\Sigma_{24}$ is the field which obtains an SU(5) breaking vev, it should be noted that the octet and triplet cannot be arbitrarily light. If their masses lie significantly below $\langle\Sigma_{24}\rangle$, one finds that $\Gamma(\Sigma_3\rightarrow hh)/m_{\Sigma_{3}}\gg1$. In the remainder of this paper we take the triplet/octet mass, $m_{38}$, to be a free parameter. It should be understood that in the case $m_{38}/\langle\Sigma_{24}\rangle\ll1$, these states are assumed to arise from an additional $\bf24$ multiplet, not associated with the breaking of SU(5)\footnote{In this case, for example, a potential of the form $m^2\Phi_{24}^2+\text{Tr}\left(\left[\Sigma_{24},\Phi_{24}\right]^2\right)$, with $\langle\Sigma_{24}\rangle=V\text{diag(2,2,2,-3,-3)}$, can give GUT scale masses to the rest of the $\Phi_{24}$ while the octet/triplet are tuned to be light with $m_{38}=m$.}.

The Yukawa Lagrangian of the model is given by\footnote{Additional terms are forbidden by the U(1)$_{PQ}$ symmetry to be discussed in \sec{sec:unification}.}
\begin{equation} \label{eq:YukawaLag}
  \mathcal{L}_Y= {\mathbf y_d}\,\overline{\Psi_{10}^c}\,H_{\overline{5}}\,\Psi_{\overline{5}} + {\mathbf y_u}\,\overline{\Psi_{10}^c}\,H_{5}\,\Psi_{10}+\sum_{a=1}^2{\mathbf Y^{(a)}}\,\overline{\Psi_{\overline{5}}^c}\,\Phi^{(a)}_{10}\Psi_{\overline{5}} \, + \,\text{h.c.}\,,
\end{equation}
where $(\Psi_{\overline{5}}+\Psi_{10})$ corresponds to a single generation of SM fermions and there is an implicit sum over generations. When considering the low energy phenomenology, we will be particularly interested in the couplings of the light leptoquark states $\boldsymbol{\Delta}_{a}$ (with $a=1,2$),
\begin{align} \label{eq:yuk-lq}
\begin{split}
  \mathcal{L}_Y & \supset \sum_{a=1}^2 Y_{ij}^{(a)}\,
  \epsilon_{\alpha\beta}\,\bar{d}_{Ri} \boldsymbol{\Delta}^\beta_{a} L^\alpha_j+\mathrm{h.c.}\\
  &= \sum_{a=1}^2 Y_{ij}^{(a)} \bar{d}_{Ri}\Big{[} \Delta_a^{(-1/3)}\nu_{Lj} - \Delta_a^{(2/3)} \ell_{Lj} \Big{]}+\mathrm{h.c.}\,,
\end{split}
\end{align}
where $\alpha$, $\beta$ are SU(2) indices, ${\mathbf Y^{(a)}}$ are two generic $3\times3$ complex matrices, and $\Delta_a$ are mass eigenstates satisfying $m_{\Delta_1}\leq m_{\Delta_2}$. In the second line we decompose the weak doublets in terms of the fields $\Delta_a^{(-1/3)}$ and $\Delta_a^{(2/3)}$, where the superscripts denote the corresponding electric charge. Notice that the matrices ${\mathbf Y^{(a)}}$ should be anti-symmetric in flavour indices if $d_{Rj}$ and $L_{Lj}$ belong to the same SU(5) multiplet $\bf\overline{5}$. However, this is not always the case as suggested by the violation of the GUT relations $m_d=m_e$ and $m_s=m_\mu$~\cite{Murayama:1991ew}.

Finally, the presence of the additional $\bf\overline{5}$ Higgs can be motivated by imposing a U(1)$_{PQ}$ symmetry, which is assumed to be spontaneously broken at intermediate scales. In addition to providing the usual DFSZ axion solution to the strong CP problem~\cite{Peccei:1977hh,*Peccei:1977ur,Zhitnitsky:1980tq,*Dine:1981rt}, this PQ symmetry plays an essential role in suppressing proton decay, as we shall show in the next section.
 
%%%%%%%%%%%%%%%%%%%%%%%%%%%%%%%%%%%%%%%%

\section{Unification and Proton Decay} \label{sec:unification}

The two-loop renormalisation group equations (RGEs) for the gauge couplings take the form~\cite{Machacek:1983tz}
\begin{equation}
\label{eq:rge}
  \mu\frac{dg_i}{d\mu}=\frac{b_i\,g_i^3}{16\pi^2}+\frac{g_i^3}{(16\pi^2)^2}\left(\sum^3_{j=1}B_{ij}\,g_j^2\right) \,,
\end{equation}
where the relevant coefficients $b_i$, $B_{ij}$, are given in Appendix~\ref{app:beta_function}. There are also two-loop contributions proportional to $g_i^3\,\text{Tr}(y^\dagger y)$. However, even in the case of the top quark Yukawa, these do not have a significant impact due to their smaller numerical coefficients. We therefore neglect them in order to avoid introducing dependence on $\tan\beta$. We use the following values for the SM parameters, defined at $\mu=m_Z$ in the $\overline{\text{MS}}$ scheme~\cite{Olive:2016xmw}:
\begin{align} \label{eq:SM_inputs}
  \alpha_3(m_Z)&=0.1181\pm0.0011 \,, \notag \\
  \alpha_\mathrm{em}^{-1}(m_Z)&=127.950\pm0.017 \,, \\
  \sin^2\theta_W(m_Z)&=0.23129\pm0.00005 \,. \notag
\end{align}

\begin{figure}[ht]
  \centering
  \includegraphics[width=0.6\textwidth]{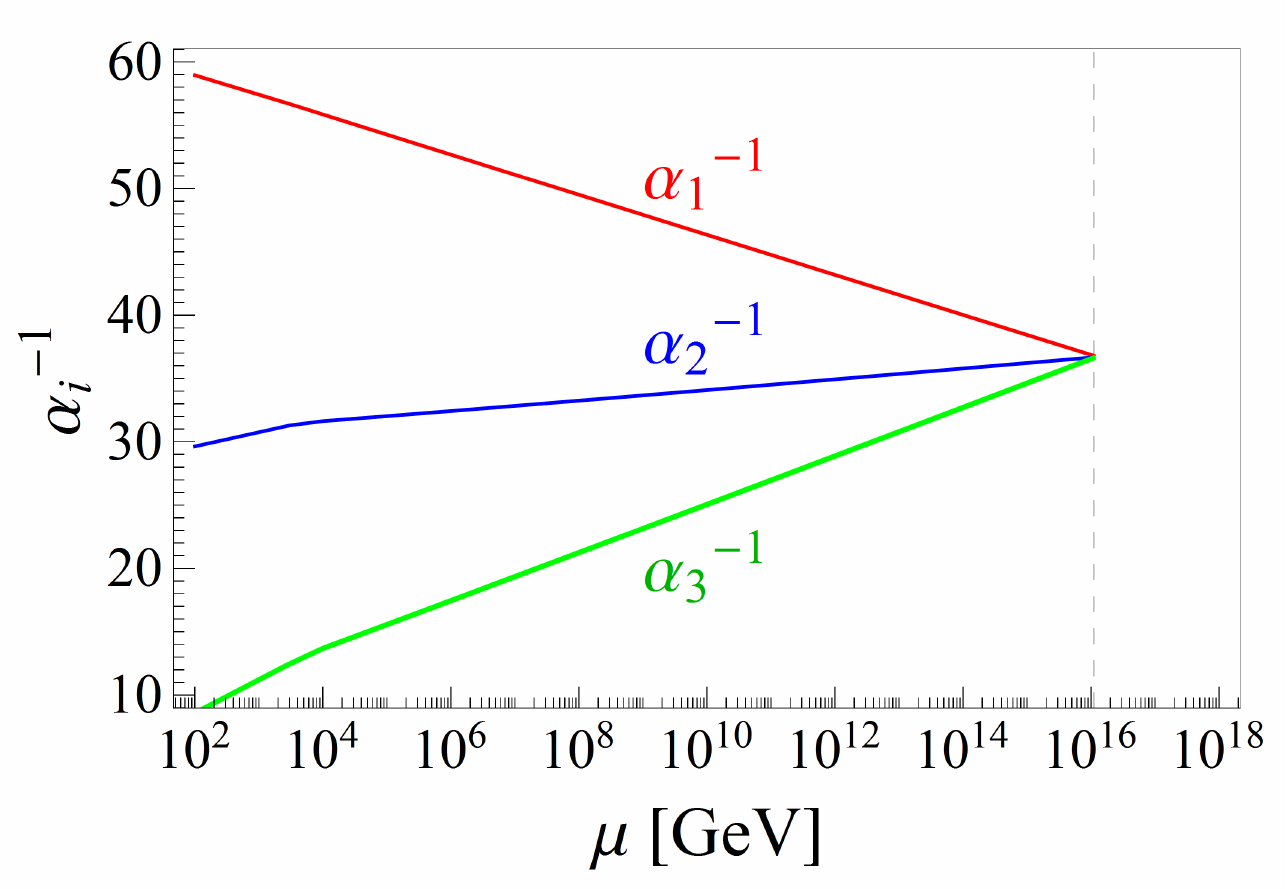}
  \caption{Running coupling constants with $m_{H}=m_{\Delta_1}=m_{\Delta_2}=3\,$TeV and $m_{38}=10~\mathrm{TeV}$. See text for details. }
  \label{fig:unification}
\end{figure}

We begin by considering the case where all scalars, including the {\bf24}-plet, can be described by complex fields, such that there are two triplet and two octet degrees of freedom. Taking the mass of the second Higgs doublet to be $m_{H}=3\,$TeV, along with the light leptoquark masses $m_{\Delta_1}=m_{\Delta_2}=3\,$TeV, and the triplet-octet mass $m_{38}=10~\mathrm{TeV}$ then leads to unification at a scale $\Lambda_{\text{GUT}}=1.2 \times 10^{16}\,$GeV, as shown in \fig{fig:unification}.

\begin{figure}[ht]
  \centering
  \includegraphics[width=0.5\textwidth]{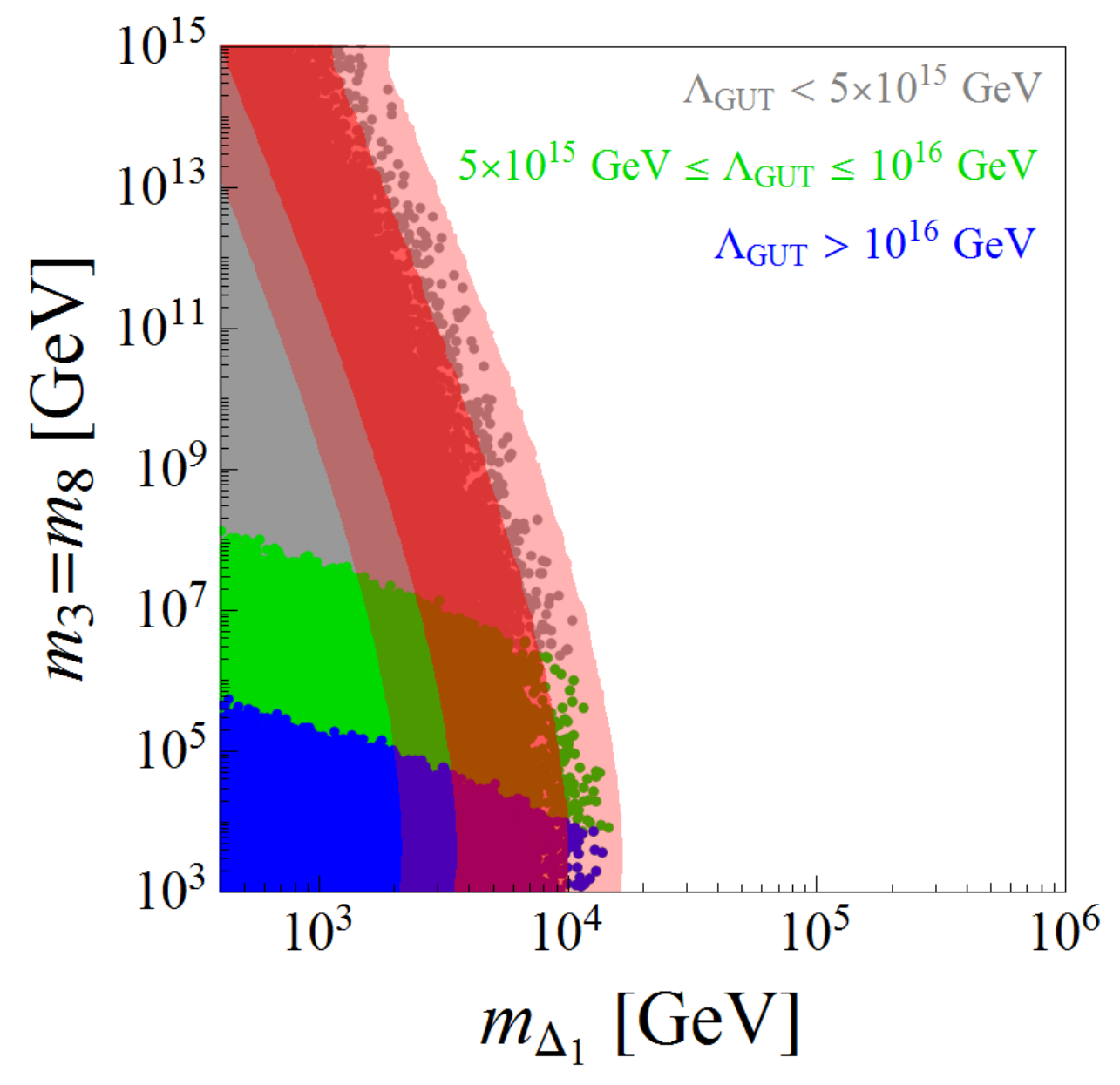}~\includegraphics[width=0.5\textwidth]{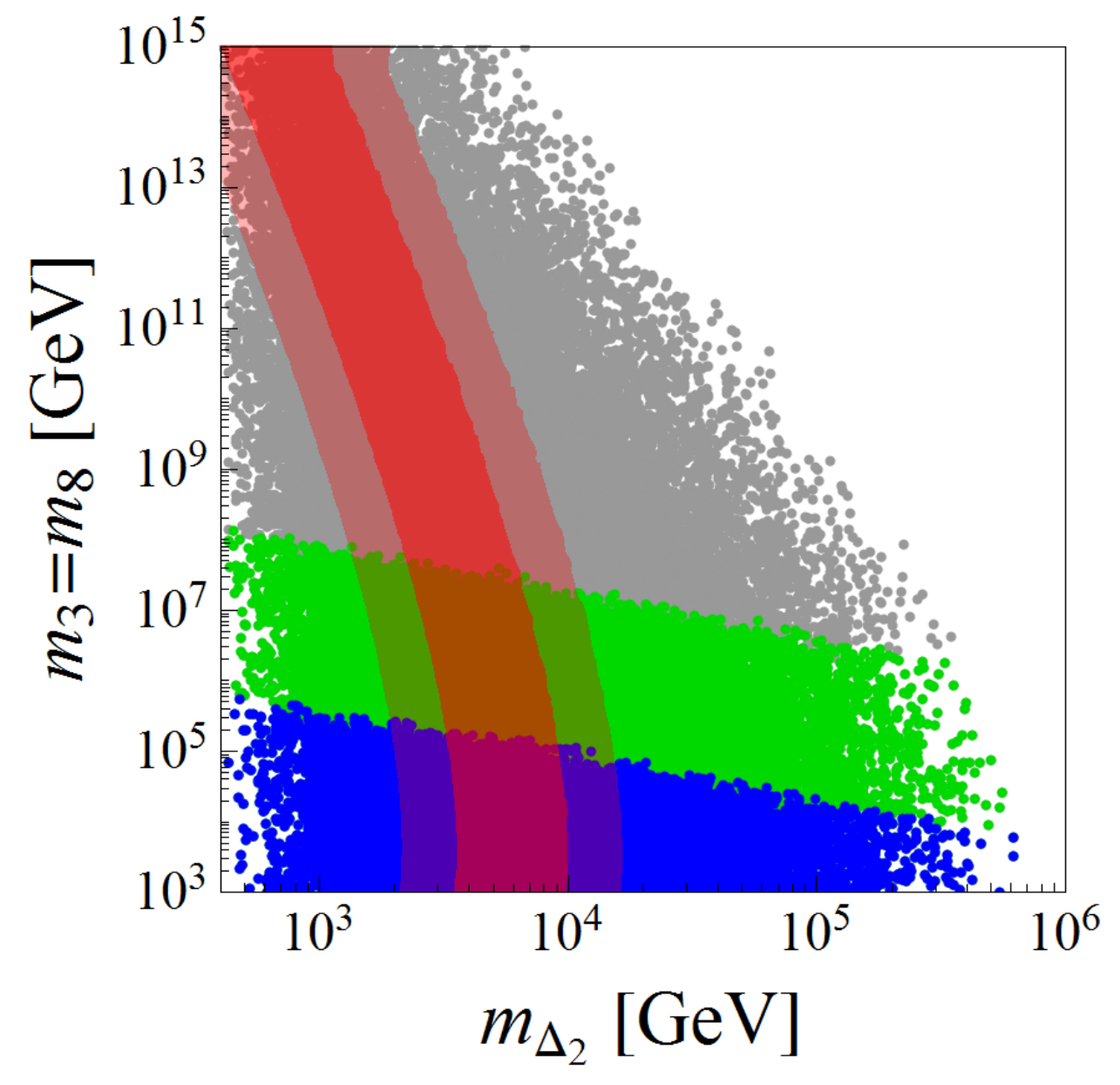}
  \caption{Regions of parameter space in which unification is obtained. The (green, blue) points correspond to a unification scale $\Lambda_{\mathrm{GUT}}>(5\times10^{15},10^{16})\,$GeV, respectively, while the grey points satisfy $\Lambda_{\mathrm{GUT}}<5\times 10^{15}\,$GeV. The dark (light) red shaded band shows the region where the couplings unify at $1\sigma$ $(2\sigma)$ in the case of degenerate leptoquark masses, $m_{\Delta_1}=m_{\Delta_2}$, and $m_{H}=480\,$GeV.}
  \label{fig:scan}
\end{figure}

To fully explore the viable parameter space of the model, we performed a random scan over the masses in the ranges
\begin{equation}
  m_{\Delta_1},\,m_{\Delta_2}\in\left[400,10^7\right]\,\text{GeV}, \quad
  m_H\in\left[480,10^7\right]\,\text{GeV}, \quad
  m_{38}\in\left[10^3,10^{16}\right]\,\text{GeV}, 
\end{equation} 
using logarithmic priors. The lower bound on the mass of the second Higgs doublet is motivated by constraints on the charged Higgs mass from $B\rightarrow X_s\gamma$~\cite{1503.01789}, while the leptoquarks are constrained by direct searches, to be discussed in \sec{sec:LHC}. The model parameters for which unification is achieved (within $2\sigma$ uncertainties on the gauge couplings) are shown in \fig{fig:scan}. From the left panel, it is evident that unification leads to an upper limit on the mass of the lightest leptoquark, $\Delta_1$. This is attained for degenerate leptoquark masses, $m_{\Delta_1}=m_{\Delta_2}$, and when $m_H$ takes its minimum value. This case is shown by the red shaded band in \fig{fig:scan}. We therefore find that unification requires at least one leptoquark to have a mass below $\lesssim16\,$TeV. Furthermore, it is clear from the right panel of \fig{fig:scan} that the second leptoquark, $\Delta_2$, cannot be arbitrarily heavy. The upper limit on $m_{\Delta_2}$ is determined by the minimal allowed values for $m_{\Delta_1}$ and $m_H$, which are constrained by experiment. Finally, note that the scale of gauge coupling unification $\Lambda_\mathrm{GUT}$, denoted by the colour of the points, is strongly correlated with $m_{38}$ and only mildly sensitive to the other mass thresholds.

Let us now comment on the minimal case, where the {\bf24}-plet is instead described by a real scalar containing a single triplet and single octet degree of freedom. Repeating the above analysis, we find results qualitatively similar to those shown in \fig{fig:scan}. However, in this case the upper bound on the lightest leptoquark mass becomes stronger, giving $m_{\Delta_1}\lesssim5\,$TeV. Perhaps more interestingly, the unification scale is now restricted to lie below $4\times10^{15}\,$GeV. While still consistent with existing bounds on the proton lifetime, this lower unification scale could potentially lead to observable proton decay at future experiments, such as Hyper Kamiokande~\cite{1109.3262}.

Even in cases where the unification scale is beyond $10^{16}\,$GeV, this model has a potentially disastrous problem due to rapid proton decay in conflict with experimental bounds. This is because in addition to the usual dimension-6 operators obtained by integrating out the $X$ and $Y$ SU(5) gauge bosons, the light leptoquarks can also mediate proton decay. This occurs via the following terms in the scalar potential\footnote{The terms in \eq{eq:scalarLag} also violate $B-L$, and hence there can be an additional constraint on $\lambda$ in order to prevent the washout of $B$ or $L$ asymmetries generated before the EWPT. This leads to the bound $\lambda^2<T/M_{Pl}$ for $T\gtrsim m_{\Delta}$ and hence $\lambda^2\lesssim m_\Delta/M_{Pl}\approx10^{-15}$.}
\begin{align} \label{eq:scalarLag}
  \mathcal{L}&\supset \lambda'\,H_{\overline{5}}\,\Phi_{10}\Phi_{10}\Phi_{10} + \lambda''\,H^\ast_{5}\,\Phi_{10}\Phi_{10}\Phi_{10} + \text{h.c.}\,,\\
   &\supset \lambda\, \epsilon_{abc}\,\Delta^a\Delta^b\Delta^cH + \text{h.c.} \,, \notag
\end{align}
where $a,b,c$ are colour indices. Proton decay then proceeds via the 5-body decay $p\rightarrow\pi^+\pi^+e^-\nu\nu$~\cite{1212.4556}. Although this decay corresponds to a dimension-9 operator, it is still problematic as the suppression scale is only $m_\Delta\sim\,$TeV. In fact, a naive estimate for the lifetime of this decay, relative to $p\rightarrow\pi^0e^+$, gives

\begin{equation} \label{eq:5body_lifetime}
\frac{\tau_{X,Y}}{\tau_\Delta}\propto\left(\frac{1}{16\pi^2}\right)^3m_p^6\left|\lambda\, Y_{11}^3\frac{v}{m_\Delta^6}\right|^2m_{X,Y}^4=4\times10^{19}\,\lambda^2\left(\frac{Y_{11}}{1/(4\pi)}\right)^6\left(\frac{1\,\text{TeV}}{m_\Delta}\right)^{12}\left(\frac{m_{X,Y}}{10^{16}\,\text{GeV}}\right)^4\,,
\end{equation}
where we have assumed loop-suppressed values for $Y_{11}$, motivated by the constraints from various low energy experiments, to be discussed in \sec{sec:flavour}. Although there is currently no dedicated search for this proton decay mode, such a short lifetime is nevertheless clearly in violation of the decay mode independent limit $\tau_p>4\times10^{23}\,$ years~\cite{nucl-ex/0104011}.

This proton decay channel was not originally identified, but was subsequently believed to be a strong reason to disfavour this model~\cite{1603.04993}. However, we wish to point out that the terms in \eq{eq:scalarLag} can be forbidden by imposing a U(1)$_{PQ}$ symmetry. Such a symmetry has additional motivation in the context of the strong CP problem and was originally considered as motivation for introducing the additional $\bf\overline{5}$ Higgs. Assigning the U(1)$_{PQ}$ charges $Q(H_5)=Q(H_{\overline{5}})=Q(\Phi^{(a)}_{10})=-2$ clearly forbids the dangerous terms. The assignment $Q=1$ for the left-handed quarks and leptons $(\Psi_{\overline{5}}+\Psi_{10})_L$, then ensures the Yukawa terms in \eq{eq:YukawaLag} are allowed by the symmetry. After PQ symmetry breaking, the terms in \eq{eq:scalarLag} will be generated by higher dimension operators, suppressed by $\Lambda_{PQ}^2/M_{\text{Pl}}^2\sim10^{-18}-10^{-12}$. Substituting this value for $\lambda$ into \eq{eq:5body_lifetime}, it is clear that the proton remains sufficiently long-lived to satisfy the existing bounds. It would however be interesting to perform a dedicated search sensitive to decays $p\rightarrow\pi^+\pi^+e^-\nu\nu$, as this could be expected to improve the current limit by several orders of magnitude. 

Of course, proton decay can still proceed via the usual dimension-6 operators and experimental bounds on the proton lifetime can then be used to place a lower bound on the unification scale. Focusing on the decay channel $p\rightarrow\pi^0 e^+$, the partial width is given by 
\begin{equation} \label{eq:decay_width}
  \Gamma\left(p\rightarrow\pi^0 e^+\right) = \frac{m_p}{8\pi}A^2\left(\frac{g_{\text{GUT}}}{\sqrt{2}\,m_{X,Y}}\right)^4 \left(\,\left| c(e^c,d) \,\langle\pi^0| (ud)_L u_R |p\rangle\right|^2 + \left| c(e,d^c) \,\langle\pi^0|(ud)_R u_L |p\rangle\right|^2\,\right) \,,
\end{equation}
where $g_{\text{GUT}}$ is the unified coupling evaluated at the mass of the $X$ and $Y$ gauge bosons, $m_{X,Y}$. The coefficients $c(e^c,d)$ and $c(e,d^c)$ depend on the fermion mixing matrices and are defined in Ref.~\cite{hep-ph/0403286}. Finally, the factor $A$ accounts for running of the four-fermion operators from $m_{X,Y}$ down to $\sim$GeV and is given by
\begin{equation} \label{eq:factorA}
  A = A_{QCD} \left(\frac{\alpha_3(m_Z)}{\alpha_3(m_\Delta)}\right)^\frac{2}{7} \left(\frac{\alpha_3(m_\Delta)}{\alpha_3(m_{38})}\right)^\frac{6}{19} \left(\frac{\alpha_3(m_{38})}{\alpha_3(m_{X,Y})}\right)^\frac{6}{16} \,,
\end{equation}
where $A_{QCD}\approx1.2$ includes the effect of running from $m_Z$ to $Q\approx2.3\,$GeV, and the light leptoquarks are assumed to be degenerate in mass. For the hadronic matrix elements we use the lattice determined values from Ref.~\cite{1304.7424}, which gives $\langle\pi^0|(ud)_R u_L |p\rangle=\langle\pi^0|(ud)_L u_R |p\rangle=0.103(41)$.

\begin{figure}[ht]
  \centering
  \includegraphics[width=0.6\textwidth]{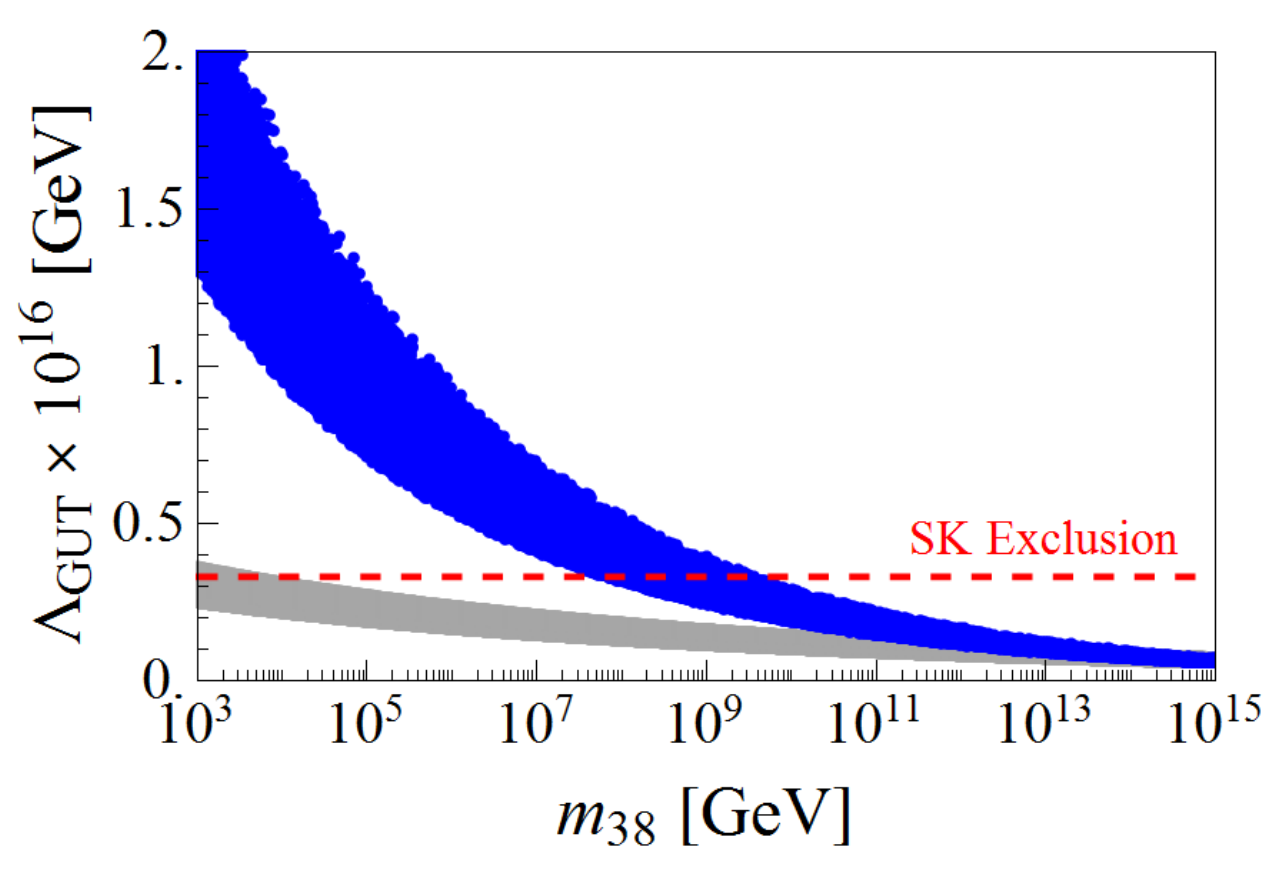}
  \caption{Unification scale as a function of the octet/triplet mass. The grey (blue) points correspond to a real (complex) octet and triplet. The dashed line shows the bound from Super Kamiokande ($\tau_p>1.29\times10^{34}$ years at 90\% CL.)~\cite{1203.4030}, assuming $c(e^c,d)=2$ and $c(e,d^c)=1$. In evaluating \eq{eq:decay_width} we have used the same masses as \fig{fig:unification}, however both the value of $g_{\text{GUT}}$ and the factor $A$ defined in \eq{eq:factorA} are approximately constant across the parameter space.}
  \label{fig:unification_SK}
\end{figure}

In \fig{fig:unification_SK} we again plot the results of the parameter scan, showing the scale of unification $\Lambda_{\text{GUT}}$, for different octet/triplet masses. The grey (blue) points correspond to the case of a real (complex) octet and triplet. Notice once again that the unification scale is only mildly dependent on the masses of the leptoquarks and second Higgs doublet. Furthermore, for octet masses close to the LHC lower bound of 1.5~TeV~\cite{ATLAS-CONF-2016-084}, the unification scale can be pushed all the way up to $\sim2\times10^{16}\,$GeV. The dashed line shows the lower bound on the unification scale derived from the Super Kamiokande limit on the proton lifetime, $\tau(p\rightarrow\pi^0e^+)>1.29\times10^{34}$ years~\cite{1203.4030}. For simplicity, we have neglected possible threshold corrections from fermions and scalars with masses near $m_{X,Y}$, such that $\Lambda_{\text{GUT}}=e^{-1/21}m_{X,Y}\simeq0.95\,m_{X,Y}$.

\fig{fig:unification_SK} suggests that, in the two cases considered, triplet/octet masses below either $\sim10^{4}\,$GeV or $\sim10^{10}\,$GeV are required in order to satisfy the bounds from Super Kamiokande. However, it should be noted that the precise bound on $m_{X,Y}$ also depends upon the details of the model at the GUT scale. This dependence is contained in the coefficients $c(e^c,d)$ and $c(e,d^c)$, which take the values $c(e^c,d)=2$ and $c(e,d^c)=1$ in the Georgi-Glashow model. These particular values also yield the maximum partial width in \eq{eq:decay_width} that is consistent with unitarity of the fermion mixing matrices. However, at the very least we know that the relation $Y_D=Y^T_E$ must be broken, which leads to some freedom in the mixing matrices. This issue was investigated in detail in~\cite{hep-ph/0410198}, where it was shown that it's possible to forbid the decay $p\rightarrow\pi^0 e^+$, along with all decays into a meson and anti-neutrinos. The leading decay mode is then into second generation fermions, $p\rightarrow K^0\mu^+$, and is suppressed by the CKM angle $\sin^2\theta_{13}$. The bound on this decay channel from Super Kamiokande is $1.6\times10^{33}\,$ years, which leads to the most conservative bound on the mass of heavy gauge bosons $m_{X,Y}\gtrsim7.4\times10^{13}\,$GeV.

%%%%%%%%%%%%%%%%%%%%%%%%%%%%%%%%%%%%%%%%

\section{$R_K$ Anomaly and Flavour Constraints} \label{sec:flavour}

The exclusive $b\to s$ transitions have been the subject of many theoretical and experimental studies over the last two decades due to their potential to probe physics beyond the SM. Recently, the LHCb collaboration measured the ratio
\begin{equation}
  R_K=\dfrac{\mathcal{B}(B^+\to K^+ \mu\mu)}{\mathcal{B}(B^+\to K^+ ee)}\Bigg{\vert}_{q^2\in[1,6]~\mathrm{GeV}^2}\,,
\end{equation}
with the dilepton invariant mass $q^2$ integrated in the $[1,6]~\mathrm{GeV}^2$ bin, and obtained \cite{Aaij:2014ora}
\begin{equation}
  R_K^\mathrm{exp}=0.745^{+0.090}_{-0.074}(\mathrm{stat})\pm 0.036(\mathrm{syst}).
\end{equation}
The result is $2.6\sigma$ smaller than the SM prediction $R_K^\mathrm{SM} \approx 1$ \cite{Hiller:2003js}. This observable is almost free of theoretical uncertainties, since hadronic uncertainties cancel out to a large extent in the ratio. Indeed, the dominant theoretical uncertainty in $R_K$ comes from QED corrections, which were estimated to be smaller than $\mathcal{O}(1\%)$ in Ref.~\cite{Bordone:2016gaq}. Therefore, if corroborated by more data, this result would be unambiguous evidence of new physics and an important hint to unveil the flavour structure beyond the SM. 

The most general dimension-six effective Hamiltonian describing the transition $b\to s\ell\ell$, with $\ell=e,\mu,\tau$, is given by \cite{Altmannshofer:2008dz}
\begin{align}
\begin{split}
  \mathcal{H}_{\mathrm{eff}} = -\dfrac{4 G_F}{\sqrt{2}}V_{tb} V_{ts}^\ast &\Bigg{\lbrace} \displaystyle\sum_{i=1}^6 C_i(\mu)\mathcal{O}_i(\mu) +\displaystyle\sum_{i=7,8} \Big{[} C_i(\mu) \mathcal{O}_i(\mu) +(C_i(\mu))^\prime (\mathcal{O}_i(\mu) )^\prime \Big{]}  \\ 
  &+\displaystyle\sum_{i=9,10,S,P} \Big{[} C_i^{\ell\ell}(\mu) \mathcal{O}^{\ell\ell}_i(\mu) +(C_i^{\ell\ell}(\mu))^\prime (\mathcal{O}^{\ell\ell}_i(\mu) )^\prime \Big{]}  \Bigg{\rbrace}+\mathrm{h.c.}\,,
\end{split}
\end{align}
where the operators relevant to our study are
\begin{align}
  \big{(}O_9^{\ell\ell}\big{)}^\prime &=\dfrac{e^2}{4\pi}\left(\bar{s}\gamma_\mu P_R b\right)(\bar{\ell}\gamma^\mu \ell)\,,\qquad\qquad
  (O_{10}^{\ell\ell})^\prime =\dfrac{e^2}{4\pi}\left(\bar{s}\gamma_\mu P_R b\right)(\bar{\ell}\gamma^\mu\gamma_5 \ell)\,.
\end{align}
The light leptoquark states in our model contribute to the following Wilson coefficients,
\begin{equation}
  \big{(}C_{9}^{\ell\ell}\big{)}^\prime = -\big{(}C_{10}^{\ell\ell}\big{)}^\prime = -\dfrac{\pi v^2}{2 V_{tb}V_{ts}^\ast \alpha_\mathrm{em} } {\displaystyle\sum_{a=1}^2} \dfrac{Y^{(a)}_{s\ell} Y^{(a)\,\ast}_{b\ell}}{m_{\Delta_a}^2}\,,
\end{equation}
where $Y_{ij}^{(a)}$ are the Yukawa couplings defined in \eq{eq:yuk-lq}. The simplest explanation of $R_K$ calls for nonzero couplings only to muons, since the couplings to first generation fermions are tightly constrained by (i) atomic parity violation experiments \cite{Dorsner:2016wpm}, (ii) the kaon physics observables and (iii) the experimental limit on $\mathcal{B}(B_s\to \mu e)$ \cite{Olive:2016xmw}. For this reason we set the couplings to the first generation to be zero, and impose the condition $Y_{b\mu}^{(a)},Y_{s\mu}^{(a)}\neq 0$ for at least one of the leptoquarks. Notice that the inequality $Y_{s\mu}^{(a)}\neq 0$ seems to be in disagreement with Eq.~\eqref{eq:YukawaLag}, which implies that the Yukawa matrices are anti-symmetric if $L_{Lj}$ and $d_{Rj}$ belong to the same $\bf\overline{5}$ multiplet. However, violation of the fermion mass relations already suggests this last assumption is not satisfied~\cite{Murayama:1991ew}, and therefore we assume that the matrices ${\mathbf Y^{(a)}}$ have a generic structure, as discussed in section~\ref{sec:model}. 

To constrain the relevant Wilson coefficient $\left(C_{9}^{\mu\mu}\right)^\prime = -\left(C_{10}^{\mu\mu}\right)^\prime$, we use the available experimental data for the $b\to s \mu\mu$ exclusive processes. Following the strategy introduced in Ref.~\cite{Becirevic:2015asa}, we perform a combined fit of $R_K$ with $\mathcal{B}(B_s\to\mu^+\mu^-)^\mathrm{exp}=\big{(}2.8^{+0.7}_{-0.6}\big{)}\times 10^{-9}$ \cite{CMS:2014xfa} and $\mathcal{B}(B\to K \mu^+\mu^-)_{\mathrm{high}~q^2}=\big{(}8.5\pm 0.3 \pm0.4\big{)}\times 10^{-8}$ \cite{Aaij:2014pli}, since the hadronic uncertainties contributing to these observables are controlled by means of numerical simulations of QCD on the lattice \cite{Aoki:2016frl}. We obtain the $2\sigma$ interval
\begin{equation}
  \left(C_{9}^{\mu\mu}\right)^\prime \in(-0.46,-0.30),
\end{equation}
where we have used the form factors computed in Ref.~\cite{Bailey:2015dka}.~\footnote{These values are consistent with the ones obtained by employing the form factors computed in Ref.~\cite{Bouchard:2013pna}.} This can be translated into the bound
\begin{equation} \label{eq:yukawa-const}
  0.031^2~\mathrm{TeV}^{-2}\leq \displaystyle\sum_{a=1,2}\dfrac{Y^{(a)}_{b\mu}Y^{(a)\ast}_{s\mu}}{{m}_{\Delta_a}^2} \leq 0.039^2~\mathrm{TeV}^{-2}\,.
\end{equation}
For sake of illustration, let us assume that only the lightest leptoquark contributes significantly to $R_K$, and impose the conservative bound $|Y_L|\lesssim 1$ to ensure that the Yukawa couplings remain perturbative all the way up to the GUT scale. Then, from \eq{eq:yukawa-const}, one can then derive the upper bound $m_{\Delta_1} \lesssim 30~\mathrm{TeV}$ on the lightest LQ mass. It is interesting to note that the unification of gauge couplings gives a stronger upper bound on the lightest LQ mass, as can be seen from \fig{fig:scan}. Therefore, if confirmed, such a violation of lepton flavour universality in meson decays can be readily obtained within our model.

Finally, regarding other low-energy signatures of the scenario discussed in this letter, a very distinctive prediction that deviates from the SM is $R_{K^\ast}=1.11(9)$ \cite{Becirevic:2015asa}, which will be tested in the near future at LHCb.~\footnote{The prediction $R_{K^\ast}>1$ is a general feature of models with $C_9^\prime=-C_{10}^\prime\neq 0$. On the other hand, scenarios in which the $R_K$ anomaly is explained through the effective coefficients $C_9$ or $C_9=-C_{10}$ predict $R_{K^\ast}<1$ \cite{Hiller:2014ula}.} In addition, depending on the LQ couplings to $\tau$s, other signatures can arise in flavour observables. For instance, lepton flavour violation in $B$ meson decays, such as $\mathcal{B}(B_s\to \mu \tau)$ and $\mathcal{B}(B\to K^{(\ast)} \mu \tau)$, can be as large as $\mathcal{O}(10^{-5})$, while the modes $\mathcal{B}(B_s\to\tau\tau)$ and $\mathcal{B}(B\to K \nu\bar{\nu})$ can be significantly enhanced with respect to the SM predictions \cite{Becirevic:2016oho}. Hence, scenarios containing the state ({\bf3},{\bf2},1/6) provide a very rich spectrum of predictions to be tested at current (LHCb) and future (Belle-II) $B$-physics experiments.

%%%%%%%%%%%%%%%%%%%%%%%%%%%%%%%%%%%%%%%%

\section{LHC Constraints} \label{sec:LHC}

The LHC collaborations have already performed a number of searches for pair-produced scalar leptoquarks using the data collected at $\sqrt{s}=13\,$TeV. Both ATLAS~\cite{1605.06035} and CMS~\cite{CMS-PAS-EXO-16-043,CMS-PAS-EXO-16-007} have released improved constraints on first and second generation leptoquarks, resulting in lower bounds on the mass of the leptoquark $m_\Delta\gtrsim1130\,(1165)\,$GeV, assuming a 100\% branching ratio $\Delta\rightarrow eq\,(\mu q)$. In the case of third generation leptoquarks, the CMS collaboration have also recently released improved bounds on leptoquarks decaying into $b\tau$~\cite{1612.01190,CMS-PAS-EXO-16-023}, obtaining a lower bound on the leptoquark mass of 900~GeV. Note that, with the exception of Ref.~\cite{CMS-PAS-EXO-16-023}, these searches currently only make use of the 2015 dataset and we can therefore expect these bounds to improve in the near future, using the significantly greater integrated luminosity collected in 2016. 

\begin{figure}[ht]
  \centering
  \includegraphics[width=0.6\textwidth]{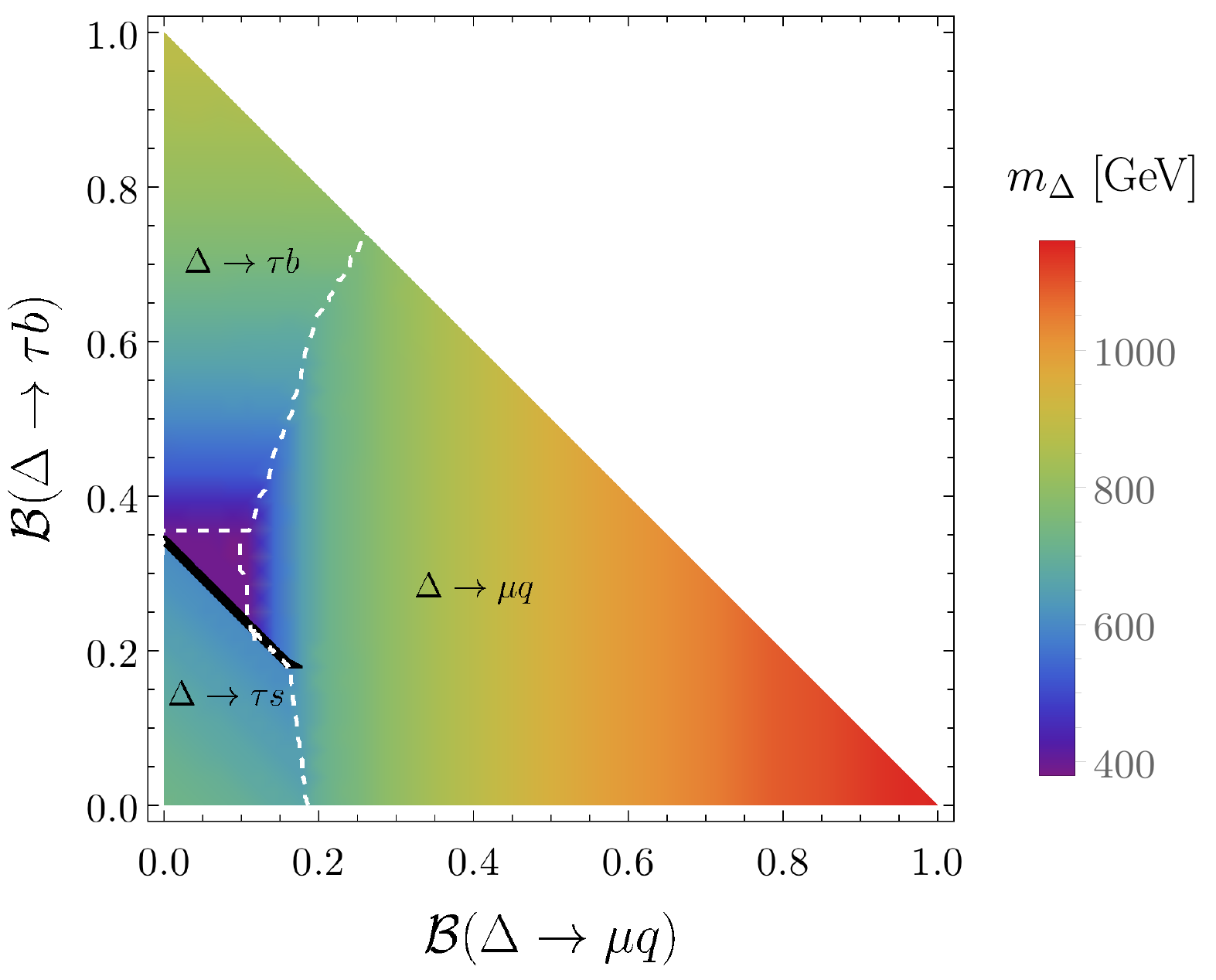}
  \caption{Lower bound on the $\Delta^{(2/3)}$ leptoquark mass using the combined results from LHC searches. We have assumed $\mathcal{B}(\Delta\rightarrow \mu q) + \mathcal{B}(\Delta\rightarrow \tau b) + \mathcal{B}(\Delta\rightarrow \tau s) = 1$. The white dashed lines separate the three regions where each of the final states yields the most stringent bound on the leptoquark mass.}
  \label{fig:LHC_limits}
\end{figure}

As  mentioned in the previous section, leptoquark couplings to the first generation fermions are already highly constrained by low energy experiments and hence we assume them to be negligible\footnote{Since LHC limits on first and second generation leptoquarks are comparable, the effect of non-zero couplings to the first generation can be approximated by additional contributions to the second generation couplings when interpreting the results in \fig{fig:LHC_limits}.}. Imposing the additional requirement that our model provides an explanation for the observed $R_K$ anomaly also necessitates $Y^{s\mu},Y^{b\mu}\neq0$ for at least one of the leptoquarks. To avoid making further assumptions on the Yukawa structure of the model, we also allow possible couplings to the third generation leptons via $Y^{s\tau}$, $Y^{b\tau}$. It's evident from \eq{eq:yuk-lq} that the $\Delta^{(1/3)}$ state will always decay into final states involving neutrinos, where there are dedicated searches in the case of the $b\nu$ final state~\cite{1508.04735}. However, we will instead focus our attention on the $\Delta^{(2/3)}$, since its couplings to charged leptons yield the strongest bounds.

The $\Delta^{(2/3)}$ has three potentially relevant decay channels, $\Delta\rightarrow\mu q$, $\Delta\rightarrow\tau b$ and $\Delta\rightarrow\tau s$, where $q=(s,b)$. As discussed above, there exist dedicated searches in the first two cases. The $\tau s$ final state, on the other hand, is significantly more challenging. Nevertheless, existing searches can still be used to constrain this case. For masses in the range 600-1000~GeV, there are bounds on resonances producing jets and hadronically decaying taus~\cite{CMS-PAS-EXO-16-016}. Below this mass range, the first and second generation leptoquark searches can be used to obtain bounds on decays to $\tau s$ in the case of leptonically decaying taus. However, the bounds are relatively weak due to the suppression by $\mathcal{B}(\tau\rightarrow e\,\nu_e\,\nu_\tau)^2\simeq0.03$. 

The constraints from LHC searches for the $\Delta^{(2/3)}$ leptoquark are combined in \fig{fig:LHC_limits}. We show the lower bound on the leptoquark mass as a function of $\mathcal{B}(\Delta\rightarrow \mu q)$ and $\mathcal{B}(\Delta\rightarrow \tau b)$, assuming $\mathcal{B}(\Delta\rightarrow \mu q) + \mathcal{B}(\Delta\rightarrow \tau b) + \mathcal{B}(\Delta\rightarrow \tau s) = 1$. Provided the branching ratio to $\mu q>60\%$, the leptoquark is constrained to have a mass above 1~TeV. However, if the leptoquark instead decays mostly via the other two decay modes, then these bounds can be drastically reduced. In that case the leptoquark could be as light as $\sim400\,$GeV and still have evaded existing searches.  

Finally, in the case where the octet and triplet have masses $\sim$TeV, they could also potentially be accessible at the LHC. The strongest bounds are from searches for pair production of the colour octet with decays into two jets, which leads to a lower limit on the mass of 1.5~TeV~\cite{ATLAS-CONF-2016-084}. This model also allows for possible decays of the octet and triplet into pairs of leptoquarks, while the triplet can also decay to two Higgs. These channels may lead to other interesting collider signatures. 

%%%%%%%%%%%%%%%%%%%%%%%%%%%%%%%%%%%%%%%%

\section{Conclusion}

A simple way to achieve SU(5) gauge coupling unification involves extending the Georgi-Glashow model to include a second $\bf\overline{5}$ Higgs and two additional scalars in the $\bf10$ representation~\cite{Murayama:1991ah}. In this paper, we have explored the parameter space of this model in detail and find that it remains consistent with the most recent experimental constraints. Unification proceeds via splitting of the $\bf10$-plets, leading to two light ({\bf3},{\bf2},1/6) scalar leptoquarks. We find that the lightest of these is required to have a mass in the range $\approx0.4-16\,$TeV. This unification scenario is therefore of particular interest as the leptoquarks may be within reach of current and/or future experiments. 

Furthermore, we have addressed two shortcomings of the original model. Firstly, we demonstrated that allowing for splitting in the $\bf24$-plet preserves unification, while providing a straightforward way to significantly raise the unification scale and hence evade the latest bounds on the proton lifetime. In the case of complex triplet and octet scalars with masses close to the current experimental lower bound of 1.5~TeV, the unification scale can be as high as $\sim2\times10^{16}\,$GeV. However, proton decay can also be mediated by the light leptoquarks, with a low suppression scale ($\sim$TeV) for the corresponding dimension-9 operator. We showed that a U(1)$_{PQ}$ symmetry can be used to forbid the dangerous terms in the Lagrangian, which would otherwise exclude this model. 

The light leptoquarks required for unification may also be directly accessible at the LHC. We have combined the results from the most recent LHC searches to derive constraints on the leptoquark parameter space. In the case where the $\Delta^{2/3}$ state decays dominantly to $\mu q$, the lower limit on its mass now exceeds 1.1~TeV. However, in the case of decays involving tau leptons, a leptoquark as light as 400~GeV is still allowed in certain regions of parameter space. A significant improvement of these results can be expected using the increased integrated luminosity collected by the LHC experiments in 2016.

Finally, there is additional motivation to revisit this model in light of recent experimental anomalies in the decays of $B$-mesons. In particular, the LHCb measurement of the theoretically clean ratio $R_K$, which shows a $2.6\sigma$ deviation from the SM prediction. It was previously shown~\cite{Becirevic:2015asa} that a ({\bf3},{\bf2},1/6) scalar leptoquark could provide a viable explanation for the anomaly. Here, we have shown that the requirements from unification are perfectly consistent with the leptoquark explanation for $R_K$. Furthermore, such leptoquarks also predict signals in other flavour observables, most notably an enhancement in the soon to be measured ratio $R_{K^\ast}$ and the possibility of lepton flavour violation in the modes $B_s\to\mu\tau$ and $B\to K^{(\ast)}\mu\tau$. If the result is confirmed with additional data, $R_K$ along with other flavour observables could therefore provide the first tentative hints towards a unification scenario similar to that considered here. 

%%%%%%%%%%%%%%%%%%%%%%%%%%%%%%%%%%%%%%%%

\section*{Acknowledgements}

O.S. would like to thank Damir Becirevic and Nejc Kosnik for useful discussions, and Kavli IPMU for its kind hospitality. A.K. and T.T.Y. thank Prof. Archil Kobakhidze for the hospitality during their stay at the University
of Sydney. This work is supported by Grant-in-Aid for Scientific Research from the Ministry of Education, Science, Sports, and Culture (MEXT), Japan,  No. 26104009 (T.T.Y.), No. 26287039 (T.T.Y.) and No. 16H02176 (T.T.Y.), and World Premier International Research Center Initiative (WPI Initiative), MEXT, Japan (P.C, A.K. and T.T.Y.). The work of A. K. was supported by the U.S. Department of Energy Grant No. DE-SC0009937. This project has received funding from the European Union's Horizon 2020 research and innovation program under the Marie Sklodowska-Curie grant agreements No 690575 and No 674896. 

%%%%%%%%%%%%%%%%%%%%%%%%%%%%%%%%%%%%%%%%

\appendix
\section{$\beta$-function coefficients} \label{app:beta_function}

In this appendix we collect the $\beta$-function coefficients appearing in Eq.~\eqref{eq:rge}. They are given above each mass threshold by
\begin{equation}
\label{eq:rge1}
  b_{i}=
  \begin{pmatrix}
    \frac{41}{10} \\
    -\frac{19}{6} \\
    -7 \\
  \end{pmatrix} 
  +\,\Theta(\mu-m_H)
    \begin{pmatrix}
    \frac{1}{10} \\
    \frac{1}{6} \\
    0
    \end{pmatrix}
  +2\Theta(\mu-m_{\Delta})
    \begin{pmatrix}
    \frac{1}{30} \\
    \frac{1}{2} \\
    \frac{1}{3} \\
  \end{pmatrix}
  +2\Theta(\mu-m_{38})
    \begin{pmatrix}
    0 \\
    \frac{1}{3} \\
    \frac{1}{2} \\
  \end{pmatrix} \,,
\end{equation}
and
\begin{align}
\label{eq:rge2}
\begin{split}
  B_{ij}=
  \begin{pmatrix}
    \frac{199}{50} & \frac{27}{10} & \frac{44}{5}\\
    \frac{9}{10} & \frac{35}{6} & 12\\
    \frac{11}{10} & \frac{9}{2} & -26 \\
  \end{pmatrix}
  +\Theta(\mu-m_H)
  \begin{pmatrix}
    \frac{9}{50} & \frac{9}{10} & 0\\
    \frac{3}{10} & \frac{13}{6} & 0\\
    0 & 0 & 0 \\
  \end{pmatrix}
  +2\,&\Theta(\mu-m_\Delta)
  \begin{pmatrix}
    \frac{1}{150} & \frac{3}{10} & \frac{8}{15}\\
    \frac{1}{10} & \frac{13}{2} & 8\\
    \frac{1}{15} & 3 & \frac{22}{3} \\
  \end{pmatrix} \\[0.4em]
  +&2\Theta(\mu-m_{38})
  \begin{pmatrix}
    0 & 0 & 0\\
    0 & \frac{28}{3} & 0\\
    0 & 0 & 21 \\
  \end{pmatrix} \,,
\end{split}
\end{align}
where $m_H$ is the mass of the second Higgs doublet, $m_\Delta$ is the mass of the two light leptoquarks and $m_{38}$ is the mass of the states ({\bf8},{\bf1},0) and ({\bf1},{\bf3},0), as defined previously. We have assumed two mass degenerate leptoquarks in Eqs.~\eqref{eq:rge1} and \eqref{eq:rge2}. The extension of these expressions to the non-degenerate case is straightforward.
%%%%%%%%%%%%%%%%%%%%%%%%%%%%%%%%%%%%%%%%

\bibliographystyle{apsrev4-1_mod}
\bibliography{SU5LQ}

%%%%%%%%%%%%%%%%%%%%%%%%%%%%%%%%%%%%%%%%
\end{document}